\documentclass[review]{elsarticle}

\usepackage{lineno}
\modulolinenumbers[5]

\journal{Nuclear Instruments and Methods in Physics Research A}

\usepackage[utf8]{inputenc}
\usepackage[english]{babel}
\usepackage[euler]{textgreek}

\usepackage{amsmath}
\usepackage{hyperref}

\usepackage{graphicx}
\usepackage{feynmp}
\usepackage{feynmp-auto}
\usepackage{wrapfig}

\usepackage{booktabs}
\usepackage{float}
\usepackage{abstract} 

\newlength{\thelinewidth}
\thelinewidth=\textwidth

\biboptions{numbers,sort&compress}

\begin{document}

\begin{frontmatter}

\title{The QUAX-g$_p$g$_s$ experiment to search for monopole-dipole Axion interaction}

%% Group authors per affiliation:
\author{N. Crescini\fnref{email}}
\fntext[email]{crescini@pd.infn.it}
\author{C. Braggio}
\author{G. Carugno}
\address{INFN, Sezione di Padova and Dipartimento di Fisica e Astronomia, Via Marzolo 8, I-35131 Padova (Italy)}

\author{P. Falferi}
\address{Istituto di Fotonica e Nanotecnologie, CNR—Fondazione Bruno Kessler, and
INFN-TIFPA, I-38123 Povo, Trento (Italy)}

\author{A. Ortolan}
\author{G. Ruoso}
\address{INFN, Laboratori Nazionali di Legnaro, Viale dell'Universit\`a 2, I-35020 Legnaro, Padova (Italy)}

\begin{abstract}
The QCD axion is an hypothetical particle introduced to solve the strong CP problem of standard model of particle physics and is of interest as a possible component of cold dark matter.
In the axion scenario, J.E. Moody and F. Wilczek showed that a new macroscopic long-range force, mediated by axion exchange, acts on electron spins, and that such force can be described in terms
of an effective magnetic field.
The QUAX-g$_p$g$_s$ experiment, carried out at INFN Laboratori Nazionali di Legnaro, is designed to search for the effects on magnetized samples of the effective field produced by unpolarized mass sources.
As this field is macroscopic, it can be detected by measuring the change of magnetization of a paramagnetic Gadolinium silicate (GSO) crystal cooled at liquid helium temperature. The axion effective field induced magnetization can be detected with a SQUID magnetometer. By varying the position of the of source masses, the induced GSO magnetization is modulated at acoustic frequencies.
Although the full QUAX-g$_p$g$_s$ sensitivity has not been yet exploited, we are able to measure a magnetization of $10^{-17}~$T at few tens of Hz. With this sensitivity we expect to further improve the upper limit of the coupling of the predicted long-range force in the $10^{-3}$ to $1~$m interval.
\end{abstract}

\end{frontmatter}

\linenumbers

\section{Introduction}
\label{intro}
In particle physics, a spontaneously broken symmetry implies the existence of a Nambu-Goldstone boson \cite{weinberg}. If the symmetry is exact, the resulting boson is always massless, and couples with fermions with a coupling constant of $g_p=m_f/F$, where $m_f$ is the mass of the fermion and $F$ is the energy scale of the spontaneously broken symmetry. On the other hand, massive bosons are produced by symmetries which are not exact, like QCD pions for the chiral symmetry, and their mass is $m_b=\Lambda^2/F$ (where $\Lambda$ is the explicit symmetry-breaking scale of the effective Lagrangian).

The search of broken symmetries at extremely high energies can be performed through the search of such exotic pseudo-Goldstone bosons, which mediates long-range ultraweak forces. The detection of a new force will account for the presence of a boson and, therefore, of a broken symmetry; the axion is a pseudo-Goldstone boson arising from the breaking of the Peccei-Quinn symmetry \cite{pq} and it could mediate forces which are dependent on its coupling.

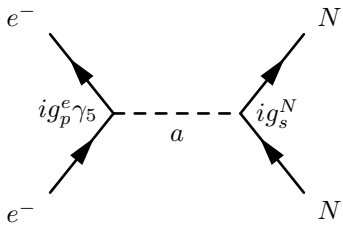
\begin{figure}[H]
\begin{center}
\begin{fmffile}{gpgs}
\begin{fmfgraph*}(120,60)
\fmfleft{i1,i2}
\fmfright{o2,o1}
\fmflabel{$e^-$}{i1}
\fmflabel{$e^-$}{i2}
\fmflabel{$N$}{o1}
\fmflabel{$N$}{o2}
\fmflabel{$ig_p^e\gamma_5$}{v1}
\fmflabel{$ig_s^N$}{v2}
\fmf{fermion}{i1,v1,i2}
\fmf{fermion}{o2,v2,o1}
\fmf{dashes,label=$a$}{v1,v2}
\end{fmfgraph*}
\end{fmffile}
\end{center}
\caption{Interaction  diagram  of a scalar-pseudoscalar coupling between a nucleus $N$ and an electron $e^-$. $N$ is unpolarized and interacts at the scalar vertex with the coupling constant $g^N_s$, whereas $e^-$ is polarized and interacts at the pseudoscalar vertex with the coupling constant $g^e_p$. The mediator is the axion $a$ and the interaction strength is proportional to $g^N_s g^e_p$.}
\label{fig1}
\end{figure}

The allowed couplings between particles is determined by the vertices of the spin-0 bosons: the pseudoscalar interaction is always spin-dependent, while, in the non-relativistic limit, the scalar interaction can be treated as spin-independent (see Fig. \ref{fig1}). Thus, in a multipole expansion, the two fields are described by the "dipole" (pseudo-scalar coupling $g_p$) and "monopole" (scalar coupling $g_s$) moments, respectively.

The different axion models (\cite{PhysRevLett.43.103,Dine1981199,Shifman1980493}) do not provide a unique value for the coupling constants $g_p$ and $g_s$. The axion mass is determined by the values of $F$, the energy scale at which the Peccei-Quinn symmetry is broken. In addiction, the presence of an axion vacuum angle expectation $\theta \neq0$ allows for monopole-dipole coupling. The values of the couplings as a function of the mass of the axion $m_a$ reads
\begin{equation}
g_p g_s=\frac{\theta \sigma}{F^2} \frac{m_u m_d}{(m_u+m_d)^2} m_a,
\label{gpgs_int}
\end{equation}
where $m_u$ and $m_d$ are the masses of the up and down quarks \cite{wilczek}.
The angle $\theta$ in the conservative Kobayashi-Maskawa model, is expected in the $10^{-14}$ range and the pion-nucleon $\sigma$ term is taken to be $60~$MeV \cite{Sainio1995}.
It is worth noticing that this value of $\theta$ is a limit from below, and that other models suggest much higher values, up to $\theta\simeq 10^{-8}$, increasing the monopole-dipole coupling of several orders of magnitude.
Cosmological and astrophysical considerations suggest a preferred value of $F\simeq10^{12}~$GeV \cite{PhysRevLett.48.1522, Preskill1983127, Abbott1983133, Dine1983137}; in this case axions would account for the missing mass of the universe. Moreover, $F$ must be $<10^{13}~$GeV, otherwise, axion would overdominate the evolution of the Universe (overclosure problem). 
The values of $\theta$, $\sigma$ and $F$ set the axion window in the coupling-mass space \cite{PhysRevD.86.015001}.

The single particle $monopole$-$dipole$ coupling (e.g. of electronic spin with nuclei) can be described by the potential
\begin{equation}
V_{md}=\frac{\hbar g_p^e g_s^N}{8 \pi m_e c} \Big[ ( \boldsymbol{\hat{\sigma}} \cdot \mathbf{\hat{r}}) \Big(\frac{1}{r\lambda_a}+\frac{1}{r^2} \Big) \Big] e^{-\frac{r}{\lambda_a}},
\label{Vmd}
\end{equation}
where $\lambda_a$ is the Compton wavelength of the axion, $g_p^e$ and $g_s^N$ are the coupling constants of the Yukawa interaction, $c$ is the speed of light in vacuum, $m_e$ is the mass of the electron, $ \boldsymbol{\hat{\sigma}}$ is the vector of Pauli spin matrices, and $r$ and $\mathbf{\hat{r}}$ are the distance and unit vector between the monopole and the dipole, respectively \cite{wilczek,weinberg,PhysRevD.52.3132}.
This potential describes the interaction between two single particles and the axion Compton wavelength determines the range of the interaction. Since $\lambda_a = h/m_a c$ is macroscopic, we have a long-range force.

Thanks to the presence of the Pauli matrices vector term, this interaction can be written in the form of an effective magnetic field. We can estimate the amplitude of this field using the formula of the potential energy of an electron in an external magnetic field $\textbf{B}$
\begin{equation}
U =\mu_e \boldsymbol{\hat{\sigma}} \cdot \mathbf{B},
\label{Um}
\end{equation}
where $\mu_e=e\hbar/2m_e$ is the Bohr's magneton and $\mu_e \boldsymbol{\sigma}$ is the spin magnetic moment of an electron. By recasting Eq.(\ref{Vmd}) as Eq.(\ref{Um}) we get the explicit expression of this field
\begin{equation}
\mathbf{B}_{\mathrm{eff,md}}=-\frac{g_p^e g_s^N}{4 \pi e c} \mathbf{\hat{r}} \Big(\frac{1}{r\lambda_a}+\frac{1}{r^2} \Big) e^{-\frac{r}{\lambda_a}}.
\label{beff}
\end{equation}
Even if the coupling between single particles is weak, a macroscopic sample, with the order of $10^{23}$ atoms, could produce a light coherent bosonic field that can be measured. From the integration of this equation over the volume of the source, we can calculate the amplitude of the equivalent measurable field. However, this field is not an ordinary magnetic field, since it couples to the spin of the fermion, and is independent  of fermion magnetic moment, electric charges, moving charges and angular momentum.
As the interaction potential is generated by pseudoscalar exchange rather than by vector gauge boson exchange, this field does not satisfy the Maxwell's equations.

The plan of the paper is as follows. In section \ref{apparatus} we describe the  experimental apparatus, consisting of source, detector and readout. Section \ref{gso_chi} is devoted to the magnetic properties of the GSO sample, section \ref{nmes} reports noise measurements, and the expected sensitivity of the experiment is calculated in section \ref{snr}  in terms of the measured sensitivity of our SQUID magnetometer. Conclusions are eventually drawn in section \ref{conclusions}.

\section{Experimental Apparatus}
\label{apparatus}

\begin{figure}[H]
\centering
\includegraphics[width=.5\textwidth]{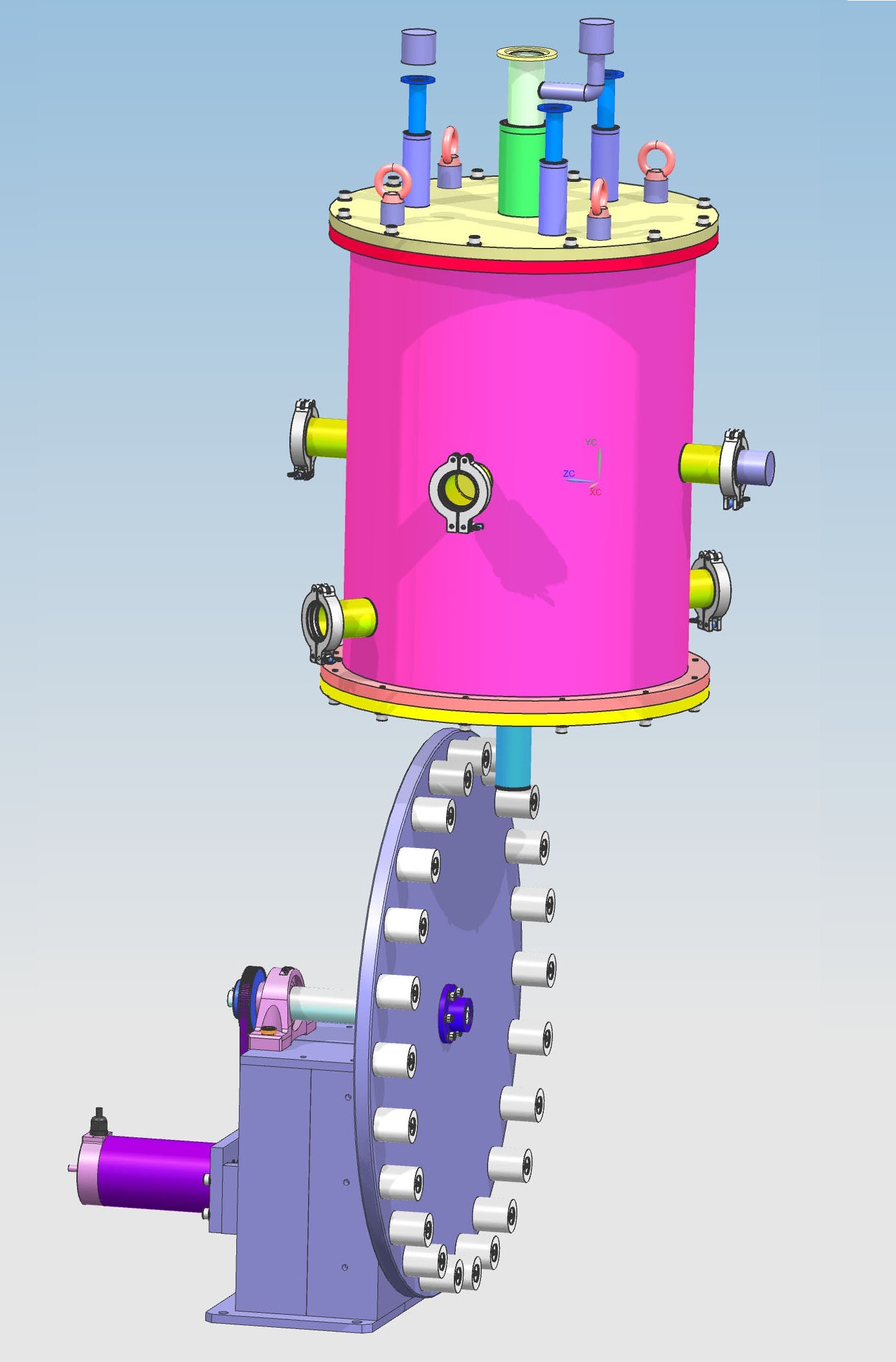}
\caption{A  schematic showing the QUAX-g$_p$-g$_s$ apparatus for measuring the strength of $g_p g_s$ interaction. The cryostat houses the GSO crystal (detector). The rotating toothed wheel (unpolarized source) modulates in time  the monopole-dipole interaction.}
\label{quax}
\end{figure}

The effective field in Eq.\ref{Vmd} interacts with electron spins of matter, and it can be detected by measuring the induced changes of  matter magnetization.
First efforts  to measure spin-dependent forces using ferromagnets and SQUIDs are reported in ref. \cite{Vorobyov1988146}. 
Our approach is similar but we substitute the ferromagnet sample with a paramagnetic crystal, in order to avoid domain wall noise. Aim of the experiment is to detect a magnetization signal which is not produced by a magnetic field but by a material with high nuclear density. The periodic modulation of the signal is achieved by varying the source-detector distance.
Figure \ref{quax} shows a schematic of the main components of the QUAX-g$_p$g$_s$ (QUest for AXion) experiment. A rotating toothed wheel allows for a source mass with a variable distance, while a cryostat houses the detector, placed as close as possible to the moving source. Hereafter we give some details about the experimental setup.
\textit{Source.} - The source consists of large unpolarized masses  that provides the monopole part of the interaction.  Each mass is a lead disk, $2.5~$cm thick, $9.5~$cm in diameter and $1.95~$kg in weight. 
In the first experimental setup, 4 masses evenly spaced are placed on a rotating wheel having controllable angular velocity; this configuration allows us to modulate the interaction at a given frequency. 
A higher frequency signal modulation is suitable to improve the apparatus sensitivity. To this aim, a measurement with 24 masses on the wheel is planned.

\textit{Detector.} - The paramagnetic sample is a cubic crystal of gadolinium oxyorthosilicate Gd$_2$SiO$_5$ (GSO) of volume $V=1~\mathrm{cm}^3$  \cite{gso}. The monopole-dipole interaction, acting on the electron spins in the crystal, causes a change in the magnetization of the sample and induces a change of the magnetic flux collected by a coil surrounding the GSO. To reduce thermal fluctuations, the crystal is  cooled down to cryogenic temperature ($T\simeq4$K). The distance between source and detector is $3.7~$cm. 

\begin{figure}[H]
\center
\includegraphics[width=.5\textwidth]{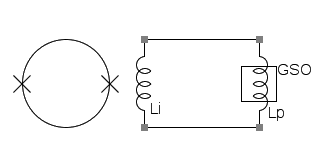}
\caption{Lumped element model of the apparatus. $L_i=1.8~\mu$H is the input coil of the SQUID, $L_p\simeq1.8~\mu$H is the pick-up coil around the GSO crystal.}
\label{circ1}
\end{figure}

\textit{SQUID readout.} - To measure the GSO magnetization induced by the source we use a Magnicon C6XXL1W dc-SQUID based magnetometer. With reference to Fig. \ref{circ1}, the coil $L_p$ picks up the magnetic flux generated by GSO magnetization; $L_p$ is connected to the SQUID input coil $L_i$. Optimal work conditions of the SQUID require $L_p\simeq L_i$. 
To further increase the sensitivity of the apparatus, the signal can be amplified with a resonant RLC circuit, tuned at the signal frequency. In the resonant readout configuration, the signal is increased by the $Q$-factor of the resonant circuit.

It is worth noticing that $B_{\mathrm{eff}}$ is not subjected to Maxwell's equations. Therefore 
we can shield the apparatus from  electromagnetic noise sources without affecting the signal. We  use two MgB$_2$ superconductive shield, that remove external noise of an expected factor $10^{12}$.  In addition, the apparatus is placed in a $\mu$-metal cylinder that reduces the magnetic fields of an expected factor of the order of $10^2$.  

\section{GSO magnetic properties} 
\label{gso_chi}
We are interested in properties of GSO crystals that affect signal and noise, i.e. magnetic susceptibility $\chi$ and spin relaxation time $\tau_M$. 
If $k_BT>>g \mu_B H$, where $H$ is the external magnetic field, the susceptibility of a set of identical ions of angular momentum $J=L+S$ is given by Curie law. When the free ions are part of a lattice in a solid structure, e.g. ions with partially filled electronic $f$-shells in insulating crystals, the generalized Curie law of paramagnetic susceptibility reads \cite{ashcroft2011solid}
\begin{equation}
\chi_{_0}(T)=\mu_0\frac{N}{V}\frac{\mu_B^2}{3}\frac{p_\mathrm{eff}^2}{k_BT},
\label{curie1}
\end{equation}
where $\mu_0$ is the vacuum magnetic permeability, $k_B$ is Boltzman's constant, $T$ is the temperature, $N/V$ is the number of ions for unit of volume, $\mu_B$ is Bohr's magneton and $p_{\mathrm{eff}}$ is the effective number of Bohr magnetons.
Here $p_{\mathrm{eff}}=g\sqrt{J(J+1)}$, where $g$ is the Land\'e factor $g(J,L,S)=\frac{3}{2}+\frac{1}{2} \Big(\frac{S(S+1) - L(L+1)}{J(J+1)} \Big)$ taking the electron $g_0$-factor to be exactly 2. 
Referring to \cite{gso}, the reported value of the effective Bohr magnetons number for a GSO crystal is $p_{\mathrm{eff}}=8.02$.
This parameter, together with the values in table \ref{tab_gso}, gives $\chi_{_0}(4\mathrm{K})=0.72$ in good agreement with reference \cite{gso}.

To calculate the thermodynamic fluctuation of the magnetization, we resort to the fluctuation-dissipation theorem \cite{landau2013statistical}, which reads
\begin{equation}
S_{M}(\omega)=\frac{2 k_BT}{\mu_0 \omega V} \mathrm{Im}(\chi(\omega)),
\label{fd}
\end{equation}
where $S_M(\omega)$ is the spectrum of the magnetization noise, and $\mathrm{Im}(\chi(\omega))$ is the imaginary part of the Fourier transform of the susceptibility.
In the Debye approximation \cite{PhysRevLett.56.1623}, and assuming $\omega \tau_M<<1$ we have
\begin{equation}
\mathrm{Im}(\chi(\omega))=\mathrm{Im}\Big(\frac{\chi_{_0}}{1+i\omega\tau_M}\Big) \simeq \chi_{_0}\omega\tau_M.
\end{equation}
Hence, the mean square fluctuation of the magnetization, integrated over a frequency band $\Delta \nu$, is
\begin{equation}
\sigma_M^2=\int_{\Delta \nu} S_M(\nu)d\nu=\frac{4k_BT \chi_{_0} \tau_M (\Delta \nu)}{\mu_0 V}.
\label{deltam}
\end{equation}
To evaluate this quantity, the spin relaxation time $\tau_M$ of GSO was measured through an EPR spectrum, obtained with a ELEXSYS 580 Bruker instrument, equipped with a dielectric cavity at $80~$K. The measurement gives a linewidth $\Delta H_{pp}=0.48~$T. Considering the gyromagnetic ratio of the electron $\gamma_e$, we get $\delta \nu=\gamma_e \Delta H_{pp}=13.5~\mathrm{GHz}$.
Since $\gamma_e / 2\pi=28.0~$GHz/T and $\tau_M=1/ (\pi\delta \nu)$, the spin relaxation time is $7.4\times 10^{-11}~$s.
\begin{table}[H]
\centering
\begin{tabular}{l l}
\toprule
Volume			& 	$V=1~\mathrm{cm}^3$	\\
Density			& 	$\rho=6.71~$g/cm$^3$		\\
Molar weight	&	$\rho_{\mathrm{mol}}=422.58~$g/mol	\\
Number of moles &	$N_{\mathrm{mol}}=0.015~$mol			\\
\midrule
Spin relaxation time	& $\tau_M=7.4\cdot10^{-11}~$s \\
Effective number of $\mu_B$	&	$p_{\mathrm{eff}}=8.02$ \\
\bottomrule
\end{tabular}
\caption{Summarized features of the used GSO crystal.}
\label{tab_gso}
\end{table}
Assuming no variation of $\tau_M$ from 80K to 4K the value of the noise depends only on the temperature; in our case $T\simeq4$K, and using the previously calculated $\chi_{_0}(4\mathrm{K})$, the level of magnetization noise is $S_B^{1/2}(\omega) = 1.2\times 10^{-16}~\mathrm{T}/\sqrt{\mathrm{Hz}}$.

\section{Magnetic noise measurements}
\label{nmes}
Referring to figure \ref{circ1}, the spectral density of the equivalent flux noise at the pickup coil of the SQUID is given by
\begin{equation}
S^{(p)}_B(\omega)=\frac{1}{(n \pi r^2)^2}\frac{(L_i+L_p)^2}{M_i^2}S_\phi(\omega),
\label{squidfinal}
\end{equation}
where $n$ and $r$ are the number of turns and the radius of the pick-up coil respectively, $L_p$ is the pick-up coil inductance, $L_i$ is the internal inductance of the SQUID, and $M_i=k\sqrt{L_s L_i}$ is the mutual inductance between $L_i$ and the loop inductance of the SQUID $L_s$. 
The intrinsic SQUID flux noise level is $S_\phi^{1/2}(\omega)\simeq0.81\mu\phi_0/\sqrt{\mathrm{Hz}}$ \cite{magnicon}, since $M_i=8.8~$nH and $\phi_0\simeq2\times10^{15}$\,Wb, using Eq.\ref{squidfinal} and calibrating the magnetic flux through our pick-up coil, the calculation yields $S_B^{(p)}(\omega)^{1/2} \simeq 7.3 \times 10^{-16}~\mathrm{T}/\sqrt{\mathrm{Hz}}$.

\begin{figure}[H]
\includegraphics[width=.5\textwidth]{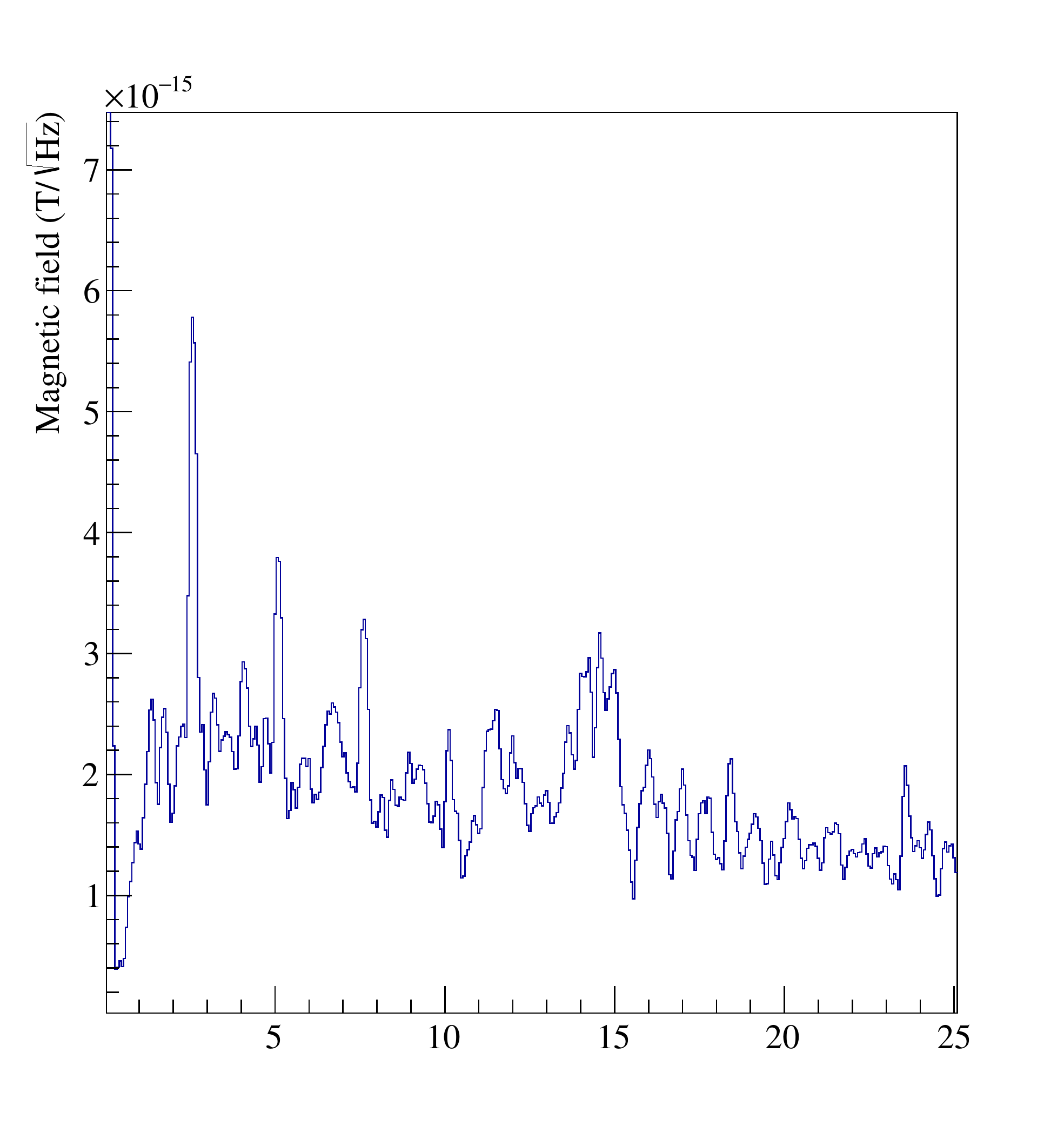} \includegraphics[width=.5\textwidth]{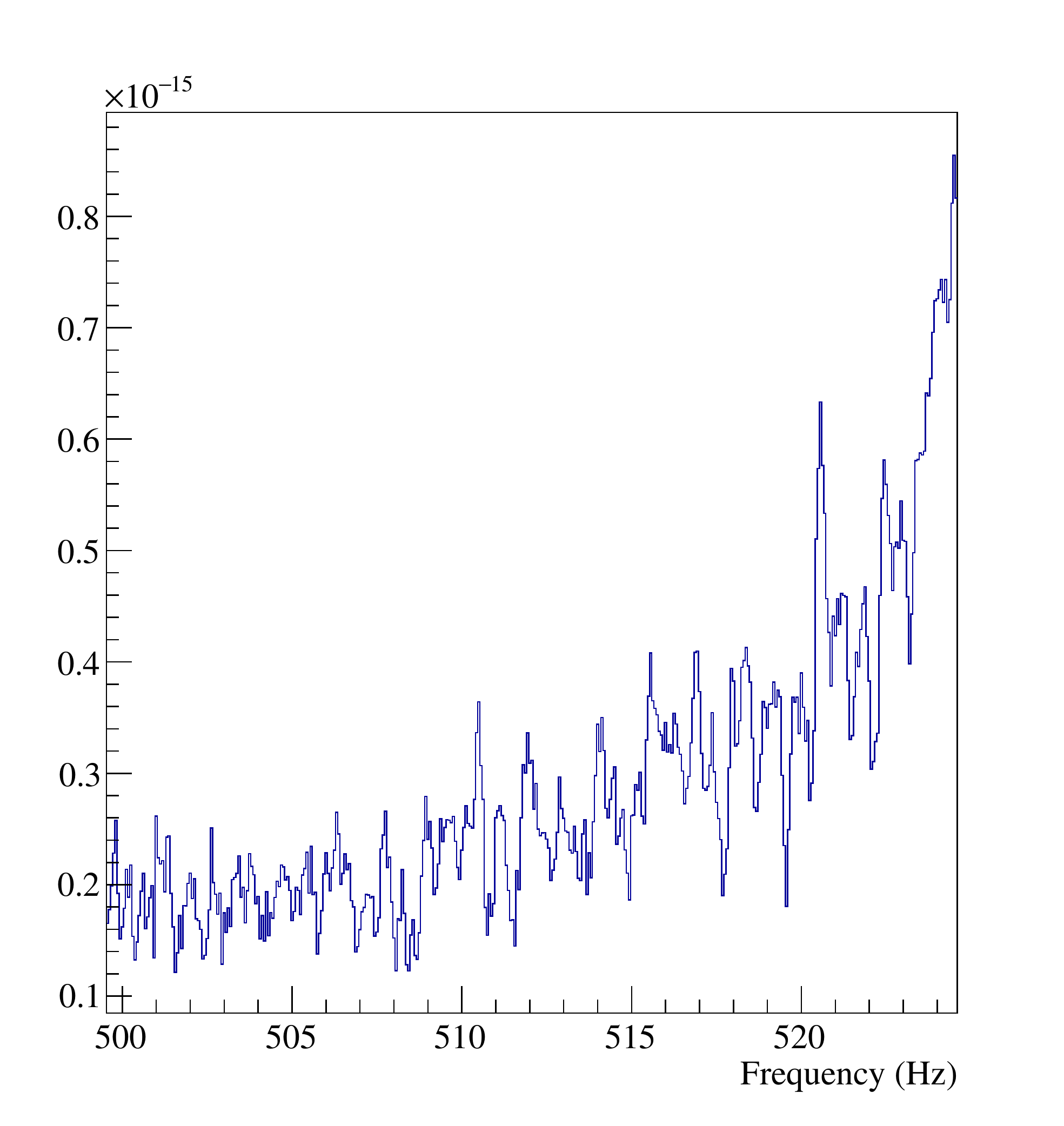}
\caption{Square root of magnetic noise power density in the frequency bands of $g_p^eg_s^N$ measurements. }
\label{spec2}
\end{figure}

Noise spectra has been measured at frequencies $f>0.1~$Hz (i.e. above the $1/f$ noise knee of the SQUID \cite{magnicon}), and are reported in figure \ref{spec2}. As can be seen, there exist frequency bands where the background is compatible with the SQUID noise.\\
The spectra are shown in the frequency bands of planned $g_p^e g_s^N$ measurements, i.e. around $10~$Hz and $500~$Hz.

\section{Sensitivity calculations}
\label{snr}
The signal that we want to measure is a change of the GSO magnetization. For a paramagnet, the change of magnetization $\mathbf{M}$ due to an external field $\mathbf{H}$ is given by $\mathbf{M}=\chi_{_0} \mathbf{H} \equiv \chi_{_0} \mathbf{B}/\mu_0$, where $\chi_{_0}$ is the magnetic susceptibility.
Likewise, the axion effective magnetic field will induce a change in the magnetization $\mu_0 M=\chi_{_0} B_{\mathrm{eff}}$, that will be collected by the pick-up coil.
The noise sources are the magnetization noise and the SQUID noise; however, in our experimental apparatus the dominant noise is due to the SQUID as discussed in sections \ref{gso_chi} and \ref{nmes}.

\textit{Non-resonant readout} - The signal to noise ratio of the magnetization measure is $\mathrm{SNR}=\chi_{_{0}} B_{\mathrm{eff}}/\sigma_B$, where $\sigma^2_B=S^{(p)}_B(\omega)/t$ is the variance of the measure and $t$ is the integration time.
The minimum detectable magnetization at unitary signal to noise ratio corresponds to an equivalent field $B_\mathrm{eff,min}=S^{(p)}_B(\omega)^{1/2}/(\chi_{_0} \sqrt{t})$. Substituting for the measured values of $\chi_{_0}$ and $S_B^{(p)}(\omega)$, and assuming an integration time $t\simeq5~$h, we get $B_\mathrm{eff,min}\simeq10^{-17}~$T.
The present configuration of QUAX-g$_p$g$_s$ should be able to improve the upper limit reported in reference \cite{ni} of one order of magnitude.

\textit{Resonant readout} - To further improve the sensitivity, we exploit a resonant pick-up circuit.
\begin{figure}[H]
\centering
\includegraphics[width=.7\textwidth]{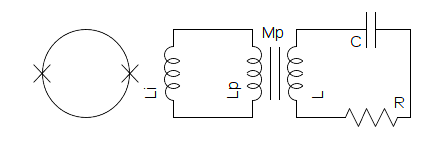}
\caption{Lumped model of the pick-up circuits coupled with the SQUID. The resistance $R$ is an equivalent resistance accounting for all circuit losses, $L$ is the inductance of the coil, $C$ is the capacity, $L_i$ is the internal inductance of the SQUID, $L_p$ is the pick-up coil, and $M_p$ is the mutual inductance between $L_p$ and $L$.}
\label{circ}
\end{figure}
Calculations show that the SNR is increased by $2Q$, with $Q$ is the quality factor of the resonant circuit. Using the measured values of the non-resonant readout and $Q\simeq10^4$, we get $B_\mathrm{eff,min}\simeq 10^{-22}~$T.

\begin{figure}[H]
\centering
\includegraphics[width=\textwidth]{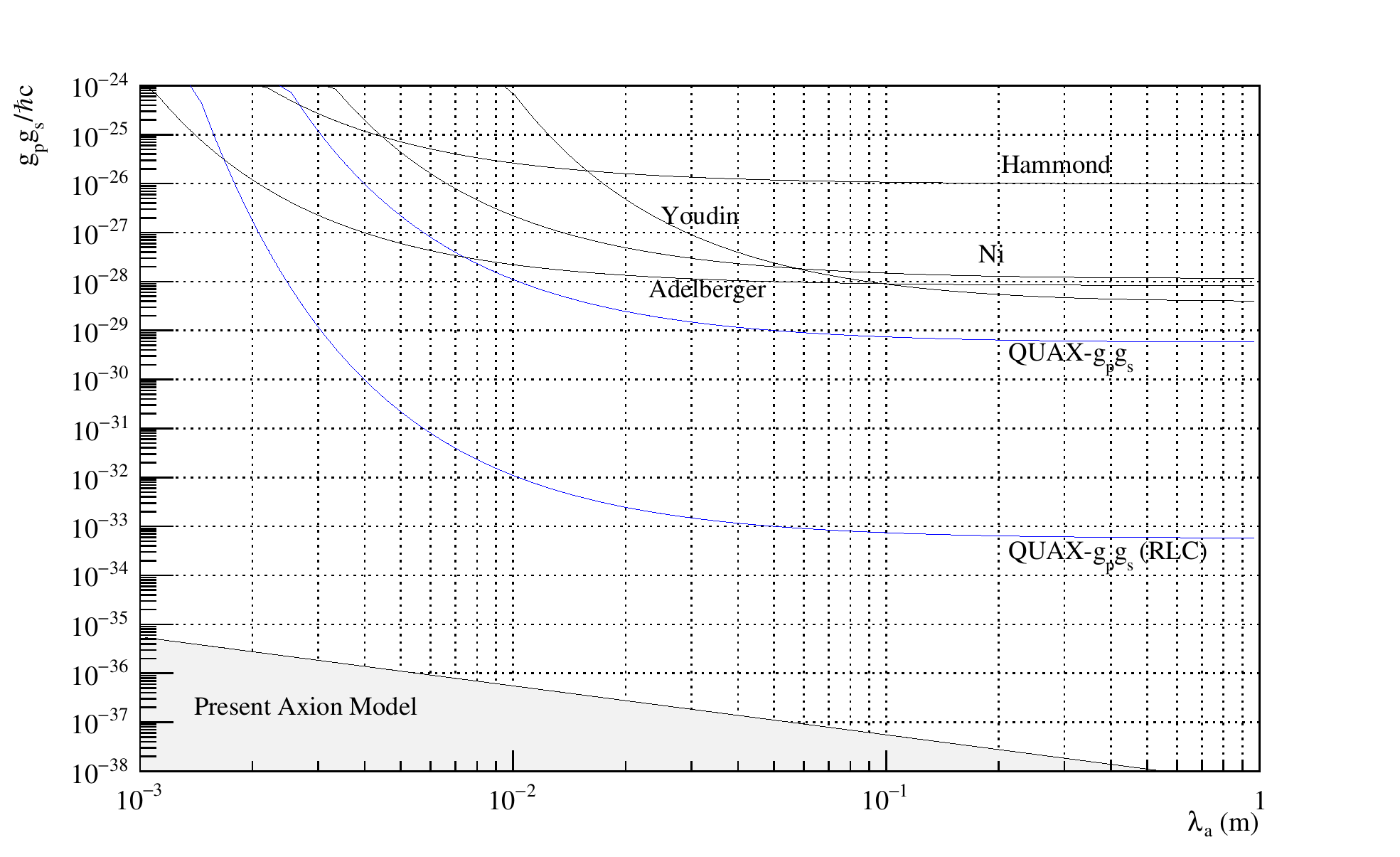}
\caption{Expected sensitivity of QUAX-g$_p$g$_s$ with an integration time of $5~$h. As can be seen the limit on the $g_p^eg_s^N$ coupling can be lowered of 1 or 5 orders of magnitude in the non-resonant or resonant(RLC) configuration, respectively. We also show the $g^e_pg^N_s$ upper limit already reported in the literature (see text).}
\label{sens_q}
\end{figure}

We summarize in figure \ref{sens_q} the expected sensitivity of our apparatus in terms of the strength of the monopole-dipole interaction, compared with the upper limits already reached by other experiments. The measurements of other authors are reported in references \citep{ni,PhysRevLett.106.041801,PhysRevLett.98.081101,PhysRevLett.82.2439,PhysRevLett.77.2170,PhysRevLett.70.701,PhysRevLett.68.135,fischbach1998search,PhysRevLett.67.1735,Daniels1994149,hsieh1989equivalence}. It is worth mentioning that some experiments has been proposed \cite{PhysRevLett.113.161801,PhysRevD.91.102006} that should be able to reach sensitivities of the same order of magnitude of our experiment.

\section{Conclusions}
\label{conclusions}
We proposed a method to measure the $g^e_p g^N_s$ interaction and reported some preliminary results on the magnetic noise of the apparatus.

Assuming an integration time of $5~$h, the minimum detectable signal is expected to be $B_\mathrm{eff,min}<10^{-17}~$T, with a resultant limit on the coupling $g^e_pg_s^N/\hbar c <10^{-29}$. 
Using a resonant circuit we should improve this upper limit of 5 orders of magnitude.

In order to reach a higher sensitivity it is possible to increase the volume of the detector. This can be done using a number $N_c$ of GSO crystals in series, read by the same SQUID with a resonant pick-up. In this configuration, to reach optimal working conditions of the SQUID one can use a superconducting low-losses matching transformer \cite{PhysRevLett.94.241101} that allows us to optimally couple $N_c$ pick-up coils to the SQUID input coil. Compared to the single crystal optimally coupled to the SQUID, this configuration would improve the SNR of $N_c^{3/2}$. 
 
\section{Acknowledgment}
We wish to thank professor Wei-Tou Ni for helpful discussions concerning the experiment. The authors also acknowledge Mario Tessaro, Enrico Berto and Fulvio Calaon for help in the realization of the experimental set-up.

\newpage
\nocite{*}
\bibliography{quax_experimental_setup}

\end{document}